\documentclass[useAMS,usenatbib,aps,floatdix,nofootinbib]{mn2e}

\usepackage{color}

\usepackage{graphics}
\usepackage{amssymb,amsmath}
\usepackage{amsmath}
\usepackage{psfrag}
\usepackage{graphicx}

\newcommand{\f}{\frac}

\newcommand{\be}{\begin{equation}}      
\newcommand{\ee}{\end{equation}}      
      
\newcommand{\bef}{\begin{figure}}      
\newcommand{\eef}{\end{figure}}      
\newcommand{\bea}{\begin{eqnarray}}    
\newcommand{\eea}{\end{eqnarray}}      
  
\def\spose#1{\hbox to 0pt{#1\hss}}
\def\ltapprox{\mathrel{\spose{\lower 3pt\hbox{$\mathchar"218$}}
\raise 2.0pt\hbox{$\mathchar"13C$}}}
\def\gtapprox{\mathrel{\spose{\lower 3pt\hbox{$\mathchar"218$}}
\raise 2.0pt\hbox{$\mathchar"13E$}}}
\def\inapprox{\mathrel{\spose{\lower 3pt\hbox{$\mathchar"218$}}
\raise 2.0pt\hbox{$\mathchar"232$}}}

\def\bse{\begin{subequations}}
\def\ese{\end{subequations}}

\def\lsim{\raise 0.4ex\hbox{$<$}\kern -0.8em\lower 0.62ex\hbox{$\sim$}} 
\def\gsim{\raise 0.4ex\hbox{$>$}\kern -0.7em\lower 0.62ex\hbox{$\sim$}}

\def\f0N{f_0^{(N)}}
\def\bec{\begin{center}}
\def\eec{\end{center}}

\title[Violent relaxation of
  ellipsoidal clouds] {Violent relaxation of
  ellipsoidal clouds}

\author[ D. Benhaiem F. Sylos Labini]{David Benhaiem${^{1}}$ and
  Francesco Sylos Labini${^{2,}}{^{1,}}{^{3}}$ \\ $^1$Istituto dei
  sistemi complessi, Consiglio Nazionale delle Ricerche, Via dei
  Taurini 19, I-00185 Rome, Italy \\ $^2$Centro Studi e
  Ricerche Enrico Fermi, Via Panisperna 00184 - Rome -
  Italy \\ $^3$INFN Unit Rome 1, Dipartimento
  di Fisica, Universit\'a di Roma Sapienza, Piazzale Aldo Moro 2,
  00185 Roma, Italy}

\begin{document}

\date{\today}

\maketitle

\begin{abstract}
An isolated, initially cold and ellipsoidal cloud of self-gravitating
particles represents a relatively simple system to { study} the
effects of the deviations from spherical symmetry in the mechanism of
violent relaxation.  Initial deviations from spherical symmetry are
shown to play a dynamical role { that is} equivalent to
{that of} density fluctuations in the case of an initially
spherical cloud. Indeed, these deviations control the amount of
particles energy change and thus determine the properties of the final
energy distribution, particularly the appearance of two species of
particles: bound and free. Ejection of mass and energy from the system
together with the formation of a density profile decaying as $\rho(r)
\sim r^{-4}$ {and a Keplerian radial velocity dispersion profile}, are
the prominent features similar to those observed after the violent
relaxation of spherical clouds.  In addition, we find that ejected
particles are characterized by highly non-spherical shapes, whose
features can be traced in the initial deviations from spherical
symmetry that are amplified during the dynamical evolution: particles
can indeed form anisotropic configurations, like bars and/or disks,
{even though the} initial cloud {was very} close to spherical.
\end{abstract}



\section{Introduction} 

The collapse of an isolated, cold and initially spherical cloud of
self-gravitating particles has been extensively studied in the
literature
\citep{Henon_1973,Van_Albada_1982,Aarseth_etal_1988,Bertin_2000,Boily_etal_2002,Joyce_etal_2009a,Syloslabini_2012a,Syloslabini_2012b}
as it is considered a paradigmatic example of the violent relaxation
mechanism.   {This is the main physical process} through which a
self-gravitating system reaches a quasi stationary virialized state
far from thermodynamic equilibrium.  Numerical experiments have shown
that the global collapse of the system occurs in a relatively short
time scale $\tau_c \sim 1/\sqrt{G \rho_0}$, where $\rho_0$ is the
cloud mass density. For $t \ltapprox \tau_c$ the violently changing
{gravitational field of the system} considerably varies
individual particle energy \citep{Henon_1964,Lynden_bell_1967}. This
mechanism of energy exchange was called violent relaxation in order to
distinguish it from the collisional two-body relaxation which occurs
on much longer time scale \citep{Binney_Tremaine_1994}.

One of the most peculiar features occurring in the collapse of a
spherical and cold cloud is represented by the ejection of a relevant
part of {its mass (i.e., $\approx 30 \%$) and energy.} While
several authors (see,
e.g.,\cite{Van_Albada_1982,David_Theuns_1989,Aguilar_Merritt_1990,Theuns_David_1990,Roy_Perez_2004,Barnes_Lanzel_Williams_2009})
noticed this ejection, the importance of this process with respect to
the mechanism of violent relaxation and to the formation of the
virialized structure has been overlooked.  In a series of works
\citep{Joyce_etal_2009a,Joyce_etal_2009b,Syloslabini_2012a,Syloslabini_2012b}
we have shown that the ejection of mass and energy is indeed the
crucial mechanism for the formation of the virialized structure
density and velocity profiles.

 Mass and energy ejection depends on both the initial shape of the
 density profile and the initial virial
 ratio~\citep{Syloslabini_2012a,Syloslabini_2012b}. Namely, the
 largest system contraction occurs for a cold and uniform initial
 density distribution; if the density profile is described by a
 power-law function of the distance from the origin, i.e. $\rho(r)
 \sim r^{-\alpha}$, the collapse gets softer when the exponent gets
 steeper.  {Eventually}, for $\alpha >2$ the system does not contract
 violently anymore and its final configuration does not substantially
 differ from the initial one \citep{Syloslabini_2012b}. On the other
 hand, when the initial velocity dispersion is large enough, i.e. $b =
 2K/W$ with $b<-1/2$ (where $K/W$ is the initial kinetic/potential
 energy), {random motions of the particles oppose to the system
   gravitational contraction,} so that the cloud only slightly changes
 its initial configuration \citep{Syloslabini_2012a}.

Only when particles ejection occurs, the density and the radial
velocity dispersion profiles of the remaining virialized structure
show the same {behavior: in particular it is observed that the
  latter scales as $\rho(r) \sim r^{-4}$ while the former scales as
  $\langle v^2_r \rangle \sim r^{-1}$}.  If ejection does not take
place then the power-law tails do not form.  This tail is indeed made
of bound particles with energy close to zero, i.e. with velocity only
slightly smaller than the escape one. The universality of the density
profile thus originates from the large spread in the energy
distribution formed during the violent collapse of the cloud.
{Because of such large energy change, some of the particles can gain
  enough kinetic energy to escape from the system while other remain
  {bound to it} but with energy very close to {zero}.}
%

Therefore a spherical and cold cloud {represents} a very simple
initial configuration showing a rich variety of non trivial dynamical
behaviors during its evolution.  {This} is however a too idealized
situation for an interesting physical initial condition.  As a further
step towards a more realistic situation we consider in this work the
collapse of a cold and isolated ellipsoidal cloud of self-gravitating
particles.  We show that the initial deviations from spherical
symmetry, finely tuned in numerical experiments, can be interpreted to
play the same role of density fluctuations for the collapse of a
spherical system, determining both the ejection of a fraction of the
system mass and the formation of the {density and radial velocity
  profiles decays respectively as $\rho(r) \sim r^{-4}$ and $\langle
  v^2_r \rangle \sim r^{-1}$}.  In addition, we show that both the
virialized structure and the ejected particles are characterized by
non-spherical shapes, whose characteristics reflect, in a different
way, the imprint of the deviations from spherical symmetry
characterizing the initial clouds.

The paper is organized as follows: we first describe in
Sect.\ref{IC-S} the initial conditions we used and the details of the
simulations we have performed.{Then in Sect.\ref{VR} we discuss
  the main features of the violent relaxation mechanism for an
  initially ellipsoidal cloud. Post-collapse virialized particles
  shows the same density and radial velocity profiles that are
  independent of the initial cloud shape: these features, together
  with the strongly anisotropic configuration of ejected particles,
  are discussed in Sect.\ref{VEP}}.  Finally we draw our main
conclusions in Sect.\ref{DC}.

\section{Initial conditions and simulations} 
\label{IC-S} 

We first describe the properties of the initial conditions 
that we considered for our controlled numerical experiments. 
Then we discuss the technicalities of the simulations.

\subsection{Initial conditions}

Initial conditions are represented by an isolated cloud of $N$
particles of equal mass $m$. The total mass of the system $M=m\cdot N$
is kept constant and thus when the number of particles $N$ is changed
their mass scales accordingly, i.e. $m = M/N$.  Note that the
thermodynamic limit corresponds to
\be
\label{TDlimit}
\lim_{N\rightarrow \infty; m \rightarrow 0}  m \cdot N = M = const. 
\ee
Particles are distributed randomly with uniform density, i.e.  density
fluctuations are Poissonian and decay as $\Delta N/ N \sim N^{-1/2}$.

The cloud has initially an ellipsoidal shape
\footnote{Our units are such that, for the spherical case
  $a_1=a_2=a_3=R_c$ we have 
$\rho_0= M/(4 \pi R_c^3/3) = 1 $ gr/cm$^3$ so that the
  characteristic time scale for the collapse (see Eq.\ref{tauc}) is
  $\tau_c=2100$ seconds}
\begin{equation}
	\frac{x^2}{a_1^2}+\frac{y^2}{a_2^2}+\frac{z^2}{a_3^2}=1 \;.
	\label{eq_ellipsoid}
\end{equation}
where we take hereafter $a_3 \ge a_2 \ge a_1$. 
The three eigenvalues of the inertia tensor can be calculated as 
\be
\label{lambda} 
\lambda_i = \frac{1}{5} M (a^2_j+a^2_k)
\ee
where $i \ne j \ne k$ and $i,j,k,=1,..,3$: from the definition of the
semi-axes we have $\lambda_1 \ge \lambda_2 \ge \lambda_3$ --- note that
${\lambda}_1$ is oriented in the direction of the smallest
semi-axis $a_1$ and ${\lambda}_3$ in the direction of the largest
one $a_3$.

To characterize the structure shape we define three different linear
combinations of $\lambda_i$: the flatness parameter\footnote{Note that
  these parameters are generally defined in function of the semi-axis
  $a_1,a_2,a_3$ and not in function of the eigenvalues (see
  Eq.\ref{lambda}). However for small deformations of a perfect
  sphere (those that we considered here) these definitions are almost
  equivalent.  The advantage of using the eigenvalues to define these
  parameters is that this choice allows one to describe more easily
  shapes that differ from ellipsoids.}
\be 
\label{iota} 
\iota = \frac{\lambda_1}{\lambda_3} -1 \;, 
\ee 
the triaxiality index
\be 
\label{tau} 
\tau = \frac{\lambda_2-\lambda_3}{\lambda_1 - \lambda_3} 
\ee
and the disk parameter 
\be 
\label{phi} 
\phi = \frac{\lambda_1-\lambda_3}{\lambda_2+ \lambda_3} \;.
\ee
Note that the combination of $\iota, \tau$ and $\phi$ allows one to
distinguish not only between different type of ellipsoids (prolate,
oblate and triaxial) but also between other shapes like bars and
disks.  In Tab.\ref{table} we report their values for some
paradigmatic cases.
\begin{table}
\begin{center}
\begin{tabular}{|c|c|c|c|}
\hline
Name           & $\iota$  & $\tau$ & $\phi$ \\ 
\hline
Sphere         &   0        &  --     & 0                \\ 
Prolate        &   $x$      &  1     & $\approx x/2$    \\ 
Oblate         &   $x$      &  0     & $\approx x/2$    \\
Triaxial       &$2x/(2+x)$  &  1/2   & $2x/(4+3x)$      \\  
Disk           &   1        &  0     & 1/2              \\
Tiny Cylinder  &   $\gg 1$  &  1     & $\approx 1$      \\
\hline
\end{tabular}
\end{center}
\caption{Oblate ($a_2 = a_3 = a_1 + x > a_1 $), prolate ($a_3 =a_2+x =
  a_1+x$) and triaxial ($a_2 = (a_1+a_3)/2$, where $a_3=a_1+x$)
  ellipsoid: in all cases we take $0< x \ll 1$.}
\label{table}
\end{table}


\subsection{Simulations} 

We have performed $N_r=10$ different simulations of the same cloud but
with different realizations of the initial Poisson noise.  { We
  have computed averages of the various physical quantities over the
  different realizations of the cloud. In addition, we considered two
  set of clouds with $N=10^4$ and $N=10^5$ particles and same total
  mass.}

{ We have used the parallel version of the publicly available
  tree-code {\tt Gadget} \citep{Springel_etal_2001,Springel_etal_2005}
  to run N-body simulations}.  We used a softening length
$\varepsilon$, i.e. the scale below which the two-body potential is
not Newtonian anymore, such that $\varepsilon/\ell(t=0) \ll 1$, where
$\ell(t=0)$ is the initial inter-particle distance.  In order to
choose a proper value for $\varepsilon$ that does not affect the
gravitational dynamics during the {\it whole} time range of the
simulations (i.e. $t < 10 \tau_c$), we have considered a necessary
(but a priori not sufficient) condition: $\varepsilon$ must be smaller
than the system size at all times.  As typical radius of the system we
took the gravitational one:
\be
\label{Rg_def} 
R_g(t)  = \frac{G M_b(t)}{|W_b(t)|}
\ee
where $M_b(t)$ and $W_b(t)$ are respectively the mass of the bound
system and its potential energy at time $t$. {We numerically
  determine the minimal value $R_g^{min} $ of $R_g(t)$ during
  {the time evolution}, that occurs for times close to the
  collapse time $\tau_c$; we thus we require that $R_{g}^{min} \gg
  \varepsilon$}.  Clearly at $t \approx \tau_c$ the inter-particle
distance is $\ell(\tau_c) \ll \varepsilon$, so that two-body
scatterings can highly differ from a pure Newtonian
situation. {However the typical time scale of two-body
  scatterings is far longer than the typical collapse time scale: for
  this reason collisional effects should not be dominant during the
  collapse phase.}

 {In order to test the effects of collisionality} we have
performed several tests by varying the smoothing length $\varepsilon$
in {\tt Gadget} (and the appropriate time-step parameters) and we have
carefully followed the collapse.  Our conclusion, that agrees with
that of \cite{Joyce_etal_2009a}, who considered the collapse of a
initially cold, spherical and uniform cloud (for which the collapse is
stronger), is that as long as we have $\varepsilon < R_{min}$,
collisionality does not play an important role in the violent collapse
phase of these systems.

{In addition, we found that the} necessary condition $\varepsilon \ll
R_{g}^{min}$ is also a sufficient one, as not only the numerical
integration conserves total energy and momentum up to a few percent
but also macroscopic quantities describing the system we are
interested in, as the density and velocity profiles, the global shape,
etc., do not show {a detectable} dependence on $\varepsilon$. In
particular, we have found a good convergence of the results for
$R_{g}^{min} / \varepsilon \in [10,200]$.  On the other hand when
$\varepsilon \ge R_{g}^{min}$ the gravitational dynamics is modified
at the scale of the system and one detects large deviations with
respect to the $\varepsilon < R_{g}^{min}$ case (see also discussion
in \cite{Joyce_etal_2009a,Syloslabini_2012a}).

Beyond the softening length, the precision of a {\tt Gadget}
simulation is determined by the internal time-step accuracy $\eta$ and
by the cell-opening accuracy parameter of the force calculation
$\alpha_F$.  We have chosen the time-step criterion 0 of {\tt Gadget}
with $\eta=0.025$.  In the force calculation we employed the (new)
{\tt Gadget} cell opening criterion with a force accuracy of $\alpha_F
= 0.001$ \citep{Springel_etal_2001,Springel_etal_2005}.


\section{Violent relaxation mechanism for an ellipsoidal cloud} 
\label{VR} 

We first review in Sect.\ref{spherical} the main elements of the
violent relaxation mechanism for initially spherical clouds with
different discretization but same total mass. Then in Sect.\ref{elli} we
discuss the case of  an initially ellipsoidal cloud considering how 
to map one problem into the other.

\subsection{The spherical case} 
\label{spherical} 

\begin{figure}
\vspace{1cm}
{
\par\centering \resizebox*{9cm}{8cm}{\includegraphics*{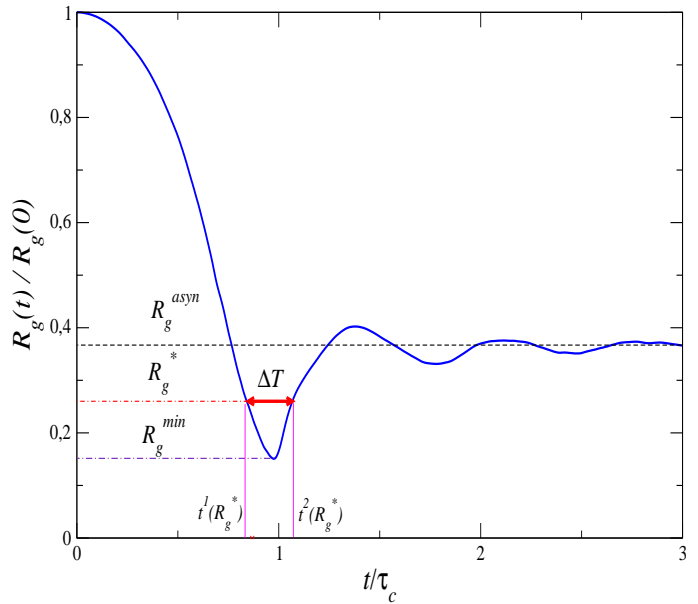}}
\par\centering
}
\caption{Example of the behavior of the gravitational radius
   {(scaled to its initial value $R_g(0)$)} as a function of time
  when the initial condition is a cold spherical cloud.  It is also
  shown the method we choose to estimate the duration of the
  collapse.}
\label{Rgrav_deltaT}
\end{figure}

For an idealized initial spherical cloud all particles have the same
collapse time, i.e. {they} arrive at the center simultaneously:
\be
\label{tauc}
 \tau_c \equiv  
\sqrt{ \frac { 3 \pi} {32 G \rho_0} }\;,
\ee
where $\rho_0 = 3 M / (4 \pi R_c^3)$. 
In the thermodynamic limit (i.e., when fluctuations can be neglected)
the collapse can be described as the simple gravitational contraction
of a perfectly homogeneous cloud. This simple idealization well
describes the observed collapse for $t < \tau_c$.  {Because of} the effect
of density fluctuations,  {the same approximation}
 fails to follow the system behavior for $t
\ge \tau_c$ \citep{Joyce_etal_2009a}, i.e. when the system size should
nominally reduce to zero to then re-expand in a periodic way.
Instead, for $t>\tau_c$ a simulated cloud relaxes into a quasi
stationary state with its size remaining almost constant (see
Fig.\ref{Rgrav_deltaT}).

\begin{figure}
\vspace{1cm}
{
\par\centering \resizebox*{9cm}{8cm}{\includegraphics*{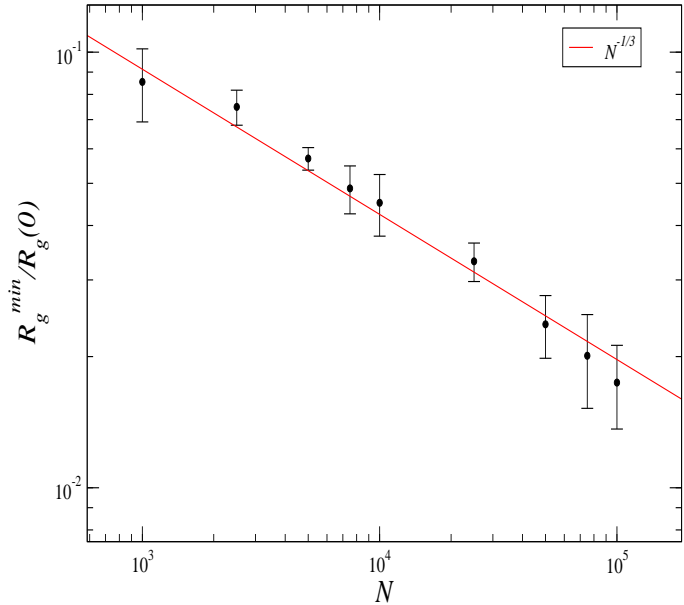}}
\par\centering
}
\caption{Gravitational radius (see Eq.\ref{Rg_scm}) for 
 an initial spherical cloud with  different values of $N$.}
\label{Rgrav_N}
\end{figure}

Several authors (see, e.g.,
\cite{Aarseth_etal_1988,Boily_etal_2002,Joyce_etal_2009a} and
references therein) found that the minimal gravitational radius of the
bound structure $R_g^{min}$ scales as
\be
\label{Rg_scm}
R_g^{min} 
\sim N^{-1/3} \;.
\ee
This relation physically means that the system stops collapsing,
reaching its minimal size, when $N$ density fluctuations on the scale
of the systems itself go nonlinear: Eq.\ref{Rg_scm} nicely fits the
observed behavior as shown in Fig.\ref{Rgrav_N}.

\begin{figure}
\vspace{1cm}
{
\par\centering \resizebox*{9cm}{8cm}{\includegraphics*{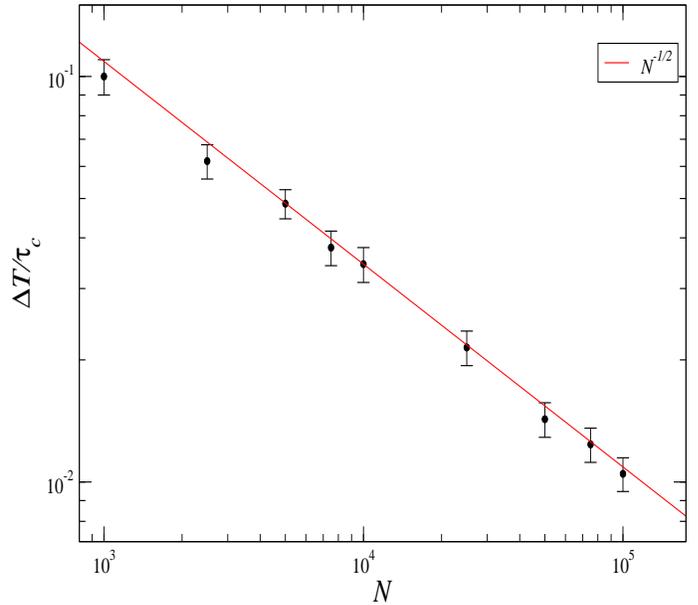}}
\par\centering
}
\caption{Behavior of the spread of free fall times $\Delta T$
  (normalized to $\tau_c$) for an initial spherical cloud and
  different values of $N$.}
\label{DeltaT_N}
\end{figure}

When $N$ is finite, the free fall time spread is related to the
initial density fluctuations.  From Eq.\ref{tauc} we simply get
\be 
\label{deltaT-scm}
\frac{\Delta T }{\tau_c} \sim \frac{\Delta \rho}{\rho_0} \sim N^{-1/2}
\;,
\ee 
which approximates well the data (see Fig.\ref{DeltaT_N}).  Note that
in the simulations $\Delta T$ has been estimated as the width of
$R_g(t)$ around its minimum \footnote{ To estimate $\Delta T$ we
  define the two values of the gravitational radius for $t < \tau_c$
  and $t > \tau_c$ such that
$R_g^*=R_g^{min} + {(R_g^{asyn} - R_g^{min})}/{2} $
where
$R_g^{asyn}$ is gravitational radius reached for $t > \tau_c$. Then
the duration of the collapse is defined as
%
$\Delta T = t^2 (R_g^*)-t^1(R_g^*) \;$ 
%
 (see Fig.\ref{Rgrav_deltaT}). }.

\begin{figure}
\vspace{1cm}
{
\par\centering \resizebox*{9cm}{8cm}{\includegraphics*{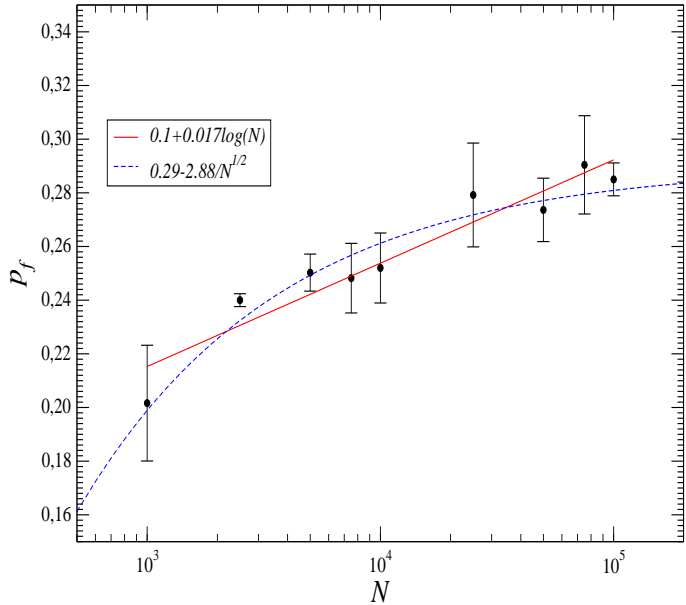}}
\par\centering
}
\caption{Fraction of ejected particles for 
different values of $N$. The best fit $p_f = 0.1 + 0.017 \log(N)$ is a
  ``blind'' numerical result, while the fit $p_f = 0.29
  -2.88/\sqrt{N}$ will be discussed in Sect.\ref{elli}.}
\label{pf_N}
\end{figure}

During the collapse a fraction of the particles may gain enough
kinetic energy to escape from the system. In particular the fraction
of ejected particles shows a slow log-dependence with $N$ (see
Fig.\ref{pf_N}) in agreement with the results of
\cite{Joyce_etal_2009a}.  {We} recall that there was no theoretical
explanation for this behavior and we will come back on this point in
Sect.\ref{elli}.
The mechanism of particles energy gain originates from the coupling of
the growth of density fluctuations with the system finite size.
Particles originally placed close to the system boundaries develop a
{net} lag with respect to the bulk, thus arriving at the system center
when the others are already re-expanding.  {These particles}, by
moving in a rapidly varying gravitational field can gain kinetic
energy and {can} drastically change the whole energy
distribution~\citep{Joyce_etal_2009a}.

The energy change particles undergo when passing through the center
has also an angular dependence that reflects the initial deviation
from perfect spherical symmetry.  Indeed, the flatness parameter
(Eq.\ref{iota}) is {initially} different from zero because of
Poisson fluctuations
\be
\label{iota0} 
\iota(0) \approx \frac{1}{\sqrt{N}} \,.
\ee
This initial small deviation from perfect spherical symmetry is
amplified during the collapse phase because of the large particles energy
change.  Eq.\ref{iota0} implies that the system is not initially
perfectly spherical, but it can be described as a ellipsoid with
$\lambda_1 > \lambda_3$ (i.e. $a_1 < a_3$). According to the energy
gain mechanism we described above, particles arriving later at the
center, i.e. those initially placed in the range $a_2 < r < a_3$, get
the largest energy kick.
Some of them are ejected but others are still bounded for $t> \tau_c$.
A simple ansatz to describe the deformation was introduced by
\cite{Worrakitpoonpon_2014,Syloslabini_2014}: if $\Delta T$ is the
time during which energy is exchanged, we can estimate that the
difference between the major and minor semi-axis is
\be
\label{iota-estimation} 
\iota(t>\tau_c) \sim \Delta T \cdot \Delta v_c \;,  
\ee 
where $v_c$ is the typical particle velocity at the collapse time.
This simply means that difference in the linear dimension between
$a_3$ and $a_1<a_3$ of the structure depends on the time interval
$\Delta T$ characterizing the collapse and on the difference in
velocity $\Delta v_c$ between particles in different directions, e.g.  along
the largest and the smallest semi-axis.  Given that the structure
is close to virial equilibrium around $t \approx \tau_c$ we have $2K_b
\approx |W_b|$ and
\be
\label{estimation_velocity} 
\Delta v_c \approx  v_c \approx \sqrt{|W_b|} \sim \sqrt{\frac{1}{R_g^{min}}} \;. 
\ee
We thus get 
\be
\label{iota-estimation2} 
\iota(t>\tau_c) \sim \frac{1}{N^{1/2}} \cdot  N^{1/6} \sim N^{-1/3}
\ee
where we used Eq.\ref{deltaT-scm}, Eq.\ref{Rg_scm} 
and Eq.\ref{estimation_velocity}.

\begin{figure}
\vspace{1cm}
{
\par\centering \resizebox*{9cm}{8cm}{\includegraphics*{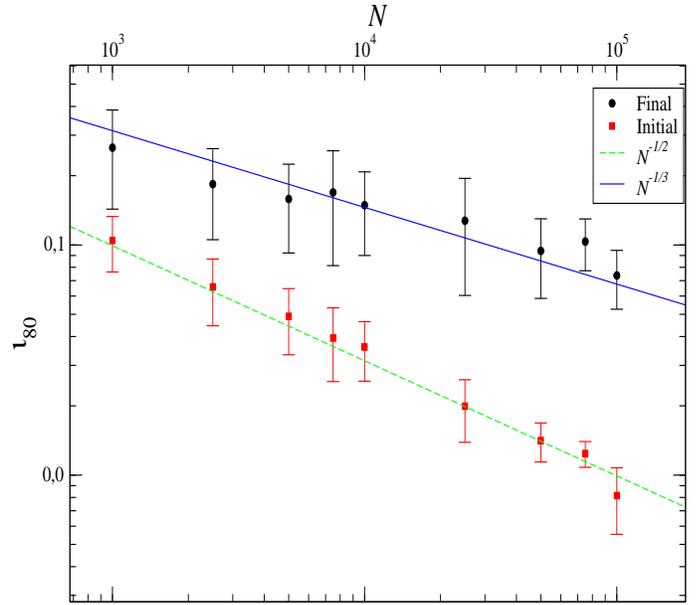}}
\par\centering
}
\caption{Behavior of $\iota_{80}(5\tau_c)$ and of the initial 
flattening ratio $\iota_{80}(0)$
  (i.e., Eq.\ref{iota0}) as a function of $N $}
\label{iota2_N}
\end{figure}

We have estimated $\iota$ for the 80 \% more bounded particles (i.e.,
$\iota_{80\%}$ --- this choice avoids the signal to be affected by
large fluctuations), at $t_{max} = 5 \tau_c$ and we considered
averages over the different realizations. As shown in
Fig.\ref{iota2_N} (where, for comparison, we report the behavior of
the initial flattening ratio $\iota_{80\%}(0)$) the agreement between
Eq.\ref{iota-estimation2} and the data is very good over two orders of
magnitude in $N$.

\begin{figure}
\vspace{1cm} { \par\centering
  \resizebox*{9cm}{8cm}{\includegraphics*{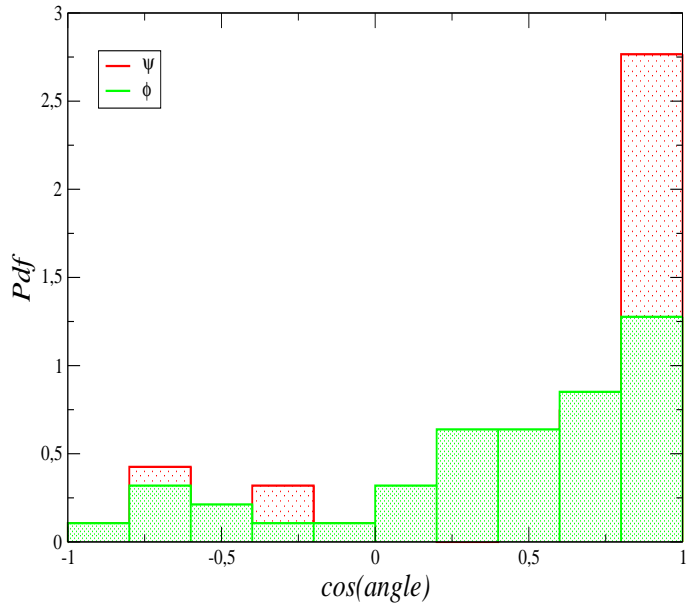}} 
\par\centering }
\caption{Probability density function (PDF) for the angles $\psi$ and
  $\varphi$ (see text).}
\label{psiphi_scm_1e4}
\end{figure}

According to the mechanism of energy gain we have discussed above, we
expect: the largest semi-axis of (i) the initial cloud at $t=0$ (ii)
the bound particles at $t_{max}$ and (iii) the free particles at
$t_{max}$ should be parallel. We call $\psi$ the angle between
(i) and (iii) and $\varphi$ the angle between (ii) and
(iii)\footnote{these are measured from the computation of the
  corresponding eigenvalue $\vec{\lambda}_3$ for bound a free
  particles at $t=,t_{max}$.}.  The results are shown in
Fig.\ref{psiphi_scm_1e4}: the probability density function for both
$\psi$ and $\varphi$ are peaked at $0$ showing that these axes are
parallel among each other.  In particular, ejection occurs more likely
in the direction of the the initial largest semi-axis, while the final
virialized structure is less correlated with such a direction.
%

%

\subsection{The ellipsoidal case} 
\label{elli} 

{When the initial cloud shape is a prolate ellipsoid then its
  initial asymmetry is amplified by the violent relaxation mechanism
  in a way that is equivalent to that of a spherical cloud. However,}
in this situation the amplitude of the deviation from spherical
symmetry, instead of being controlled by Poisson fluctuations (i.e.,
Eq.\ref{iota0}) is {tuned by the ratio between the largest $a_3$
  and the smallest $a_2=a_1$ semi-axis} \footnote{{We consider a
    situation in which the initial flatness ratio is larger than
    Poisson fluctuations, i.e. $\iota(0) \gg N^{-1/2}$}}.
%

\begin{figure}
\vspace{1cm} { \par\centering
  \resizebox*{9cm}{8cm}{\includegraphics*{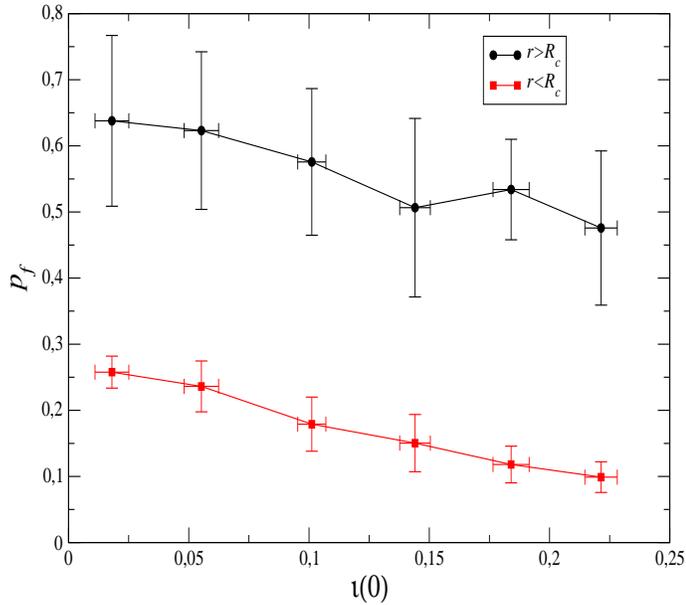}}
  \par\centering }
\caption{Black (red) dots show the fraction of particles initially
  with $r > R_c$ ($r<R_c$) that are ejected after the collapse
  (average over 20 realizations).}
	\label{ejected_fraction_ell}
\end{figure}
\begin{figure*}
\vspace{1cm} { \par\centering
  \resizebox*{5cm}{6cm}{\includegraphics*{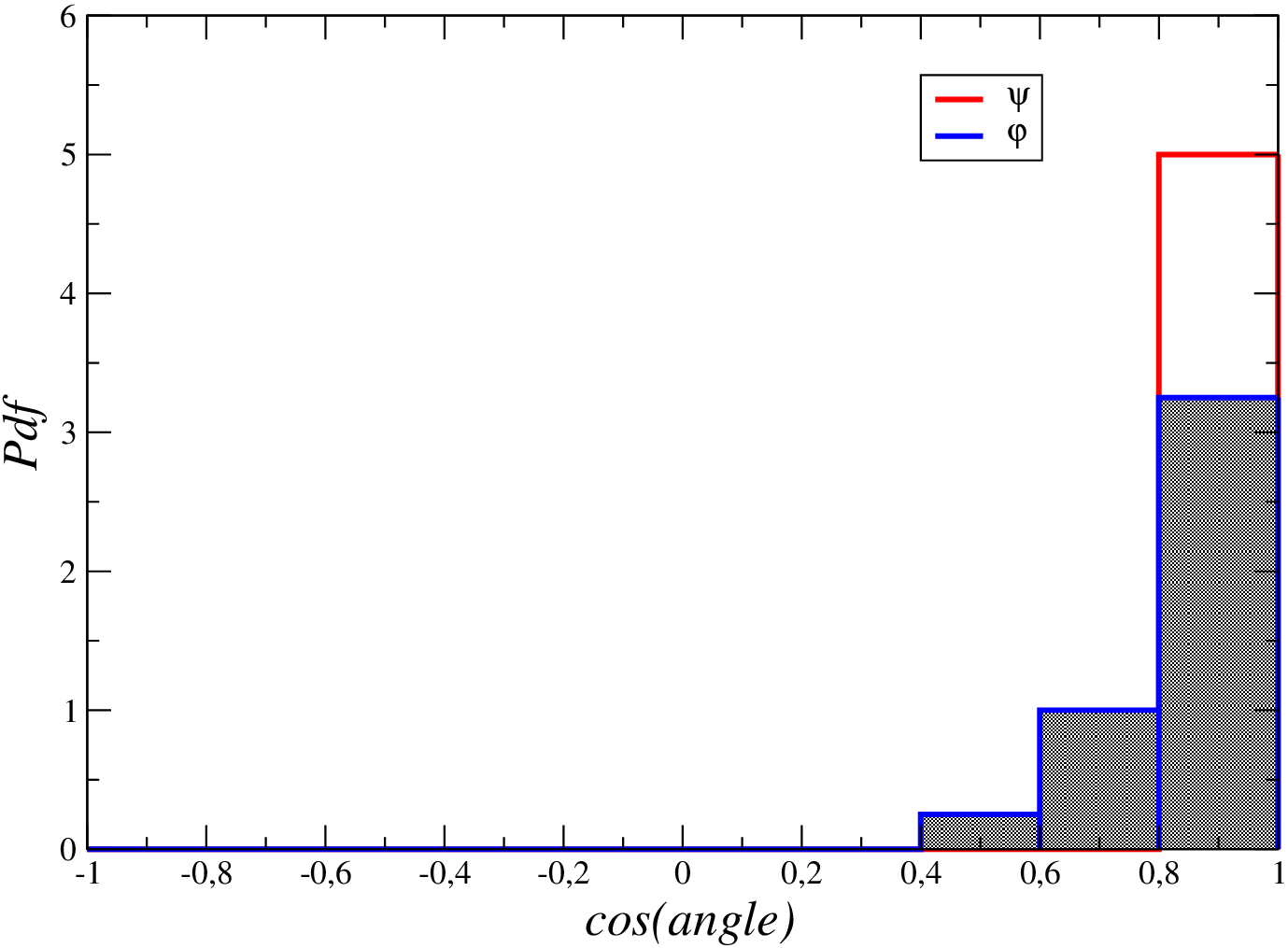}}
  \resizebox*{5cm}{6cm}{\includegraphics*{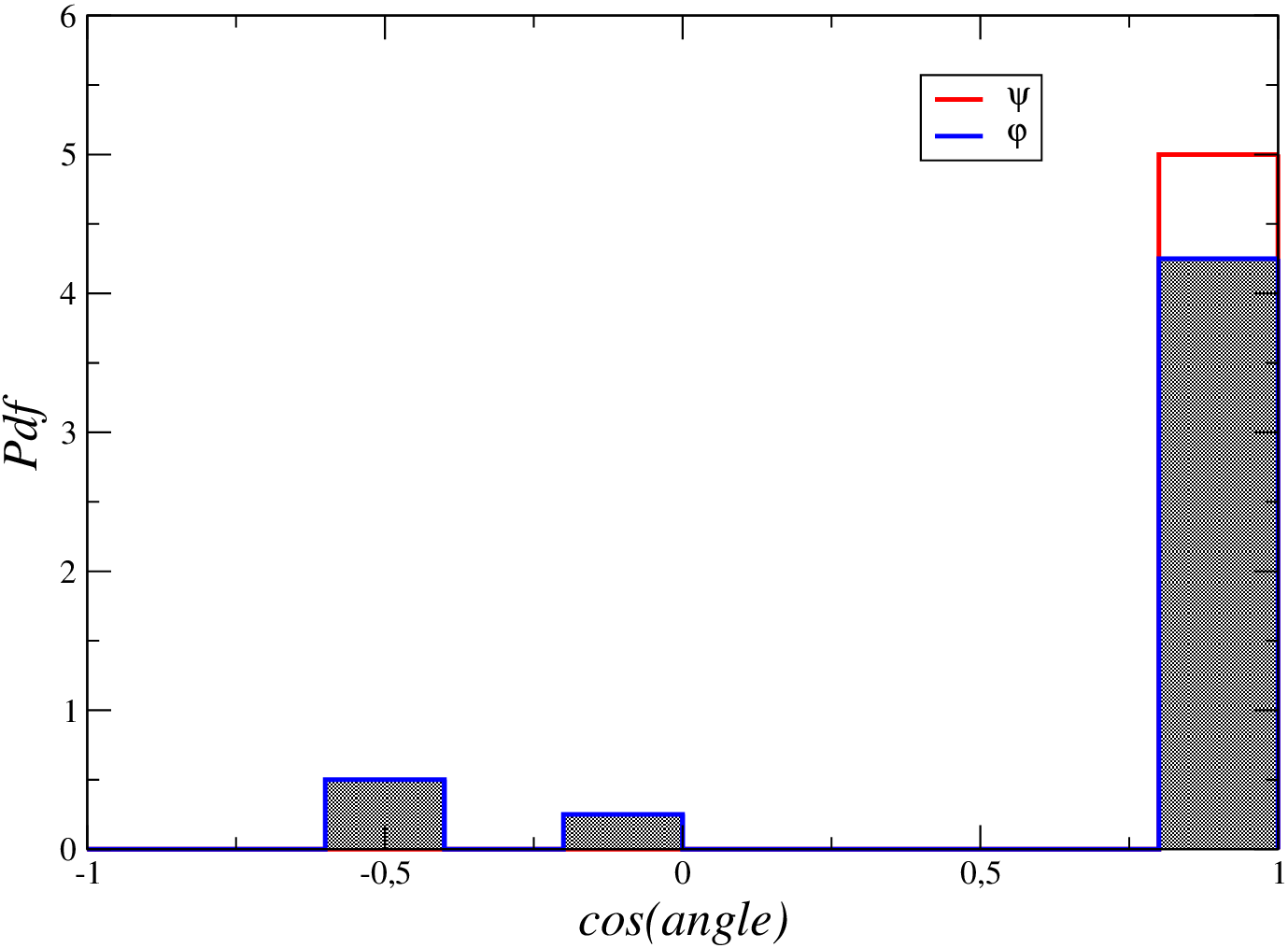}}
  \resizebox*{5cm}{6cm}{\includegraphics*{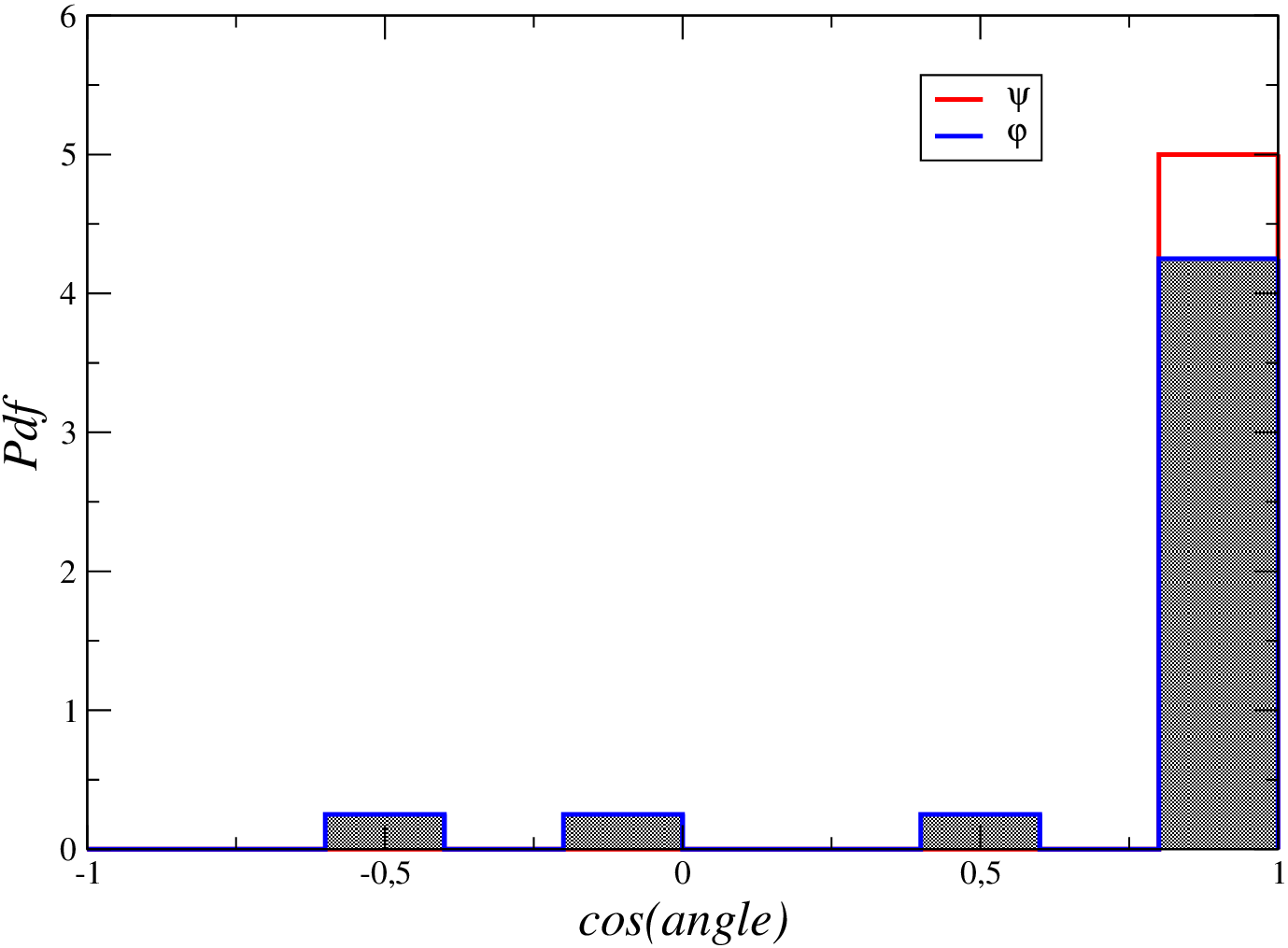}}
  \par\centering }
\caption{ As Fig.\ref{psiphi_scm_1e4} but for simulations with initial
  flattening ratio $\iota(0)=0.05, 0.15, 0.25$ (from left to right).
} \label{psiphi_ell_1e4}
\end{figure*}

Indeed, we find that, as for spherical case, ejected particles are
mostly those which originally lie close to the boundaries of the
system:  {i.e. particles} for which initially
$r_0>R_c=a_1$ (see Fig.\ref{ejected_fraction_ell}).
In addition, Fig.\ref{psiphi_ell_1e4} shows the histogram of the
values of $\psi,\varphi$ (see Sect.\ref{spherical} for definitions) for
three different sets of simulations with different initial
$\iota(0)$. The result is similar to the spherical case (see
Fig.\ref{psiphi_scm_1e4}): the largest semi-axis of the initial
conditions and of the post collapse bound and free particles are
parallel among each other.


According to the model of ejection and system deformation discussed
for the spherical case, {the increase of $\iota(0)$ corresponds
  to a decrease in the number of {particles for an initially
    spherical cloud} (see Eq.\ref{iota0}) so that: (i) the spread in
  arrival time $\Delta T$ increases and the collapse becomes softer as
  particles arrive at the center at different times, (ii) the fraction
  of ejected particles decreases and (iii) the minimum gravitational
  radius {reached during time evolution} becomes larger}:
\begin{itemize} 


\item Fig.\ref{fig-DeltaT_vs_iota0} shows that the spread in the arrival
time $\Delta T$ increases linearly with $\iota_{80}(0)$.  This can be
easily explained by considering that the spread of arrival time
$\Delta T$ is equal to the difference between the free fall time
respectively of a particle initially placed at $R_c(1+ \iota(0))$ and
of a particle initially placed at $R_c$ (for small enough $\iota(0)$):
\begin{equation}
\Delta T \approx \sqrt{\frac{3 \pi^2}{8 G \rho_0}} \left(
\sqrt{\left(1+\iota(0)\right)^3} -1 \right) \approx \frac{3}{2} \tau_c 
\iota(0) \;.
\label{DeltaT_estimation}
\end{equation}
(For the spherical cloud we used Eq.\ref{deltaT-scm} and
Eq.\ref{iota0} to compute $\Delta T$ in function of $\iota(0)$.)  
\begin{figure}
\vspace{1cm} { \par\centering
  \resizebox*{9cm}{8cm}{\includegraphics*{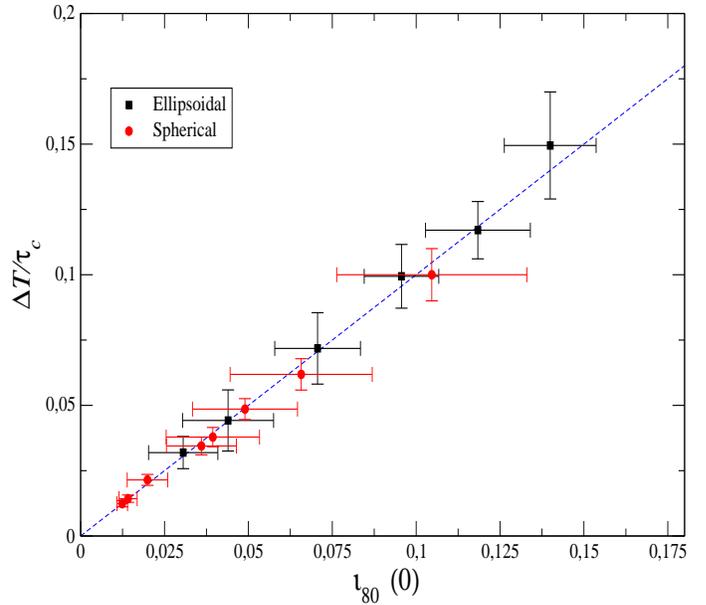}}
  \par\centering }
\caption{Average over 20 realizations of the spread in the arrival
  time $\Delta T$ versus the initial flattening ratio $\iota(0)$ for
  the case of ellipsoidal initial conditions (black dots). In addition
  we plot $\Delta T$ versus the actual initial flattening ratio for
  the simulations of a purely spherical cloud with different $N$ (red
  dots), i.e. $\iota(0) \sim N^{-1/2}$ --- see Fig.\ref{iota2_N}.}
	\label{fig-DeltaT_vs_iota0}
\end{figure}

\item Fig.\ref{fig-P_f} shows the fraction of ejected particles as a
function of $\iota_{80}(0)$ for both the ellipsoidal and the spherical
clouds: $p_f$ decreases with $\iota(0)$ as the collapse gets
softer. In particular, we note that $p_f \propto \iota(0)$ for both the
ellipsoidal case at fixed $N$ and varying $\iota(0) > N^{-1/2}$ and
for the spherical case with varying $N$ and thus with $\iota(0) \sim
N^{-1/2}$.  This result suggests a straightforward interpretation of
the fraction of mass ejected in function of the number of particles
$N$ for the spherical case (see Fig.\ref{pf_N}) in terms of
finite-size $N$-dependent deformation of the initial spherical cloud.
In addition, if we extrapolate the behavior for $\iota(0) \rightarrow
0$, which for the spherical case corresponds to $N \rightarrow \infty$,
we find that there is a saturation in the fraction of mass
ejected \footnote{For a discussion about the Vlasov-Poisson limit of
  this system see \cite{Joyce_etal_2009a}}.
%
\begin{figure}
\vspace{1cm} { \par\centering
  \resizebox*{9cm}{8cm}{\includegraphics*{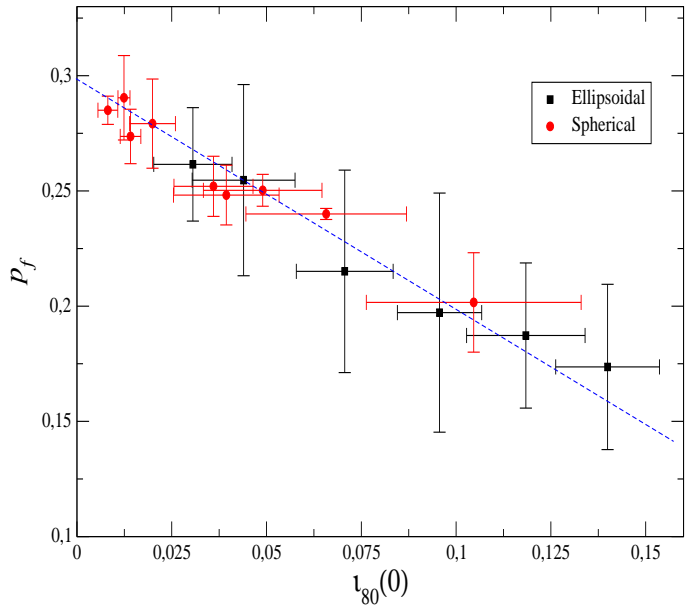}}
  \par\centering }
	\caption{Average over 20 realizations of the fraction of
          ejected particles vs $\iota_{80}(0)$ (black dots). The case
          of a uniform sphere with different $N$ is also shown (red
          dots). The dashed line corresponds to the fit $p_f \sim
          1/\sqrt{N} + const.$ in Fig.\ref{pf_N}. }
	\label{fig-P_f}
\end{figure}

\item Fig.\ref{fig-R_min} reports the minimum value of the gravitational
radius during time evolution $R^{g}_{min}$ as a function of $\iota(0)$
for both spherical and ellipsoidal initial conditions, with different
$N$ and $\iota(0)$ respectively, together with the theoretical fit:
\begin{equation}
  \label{Rg-iota}
	R_{min}^g \sim \iota(0)^{2/3}
\end{equation}
where we used, for the spherical case, Eq. \ref{Rg_scm} and
Eq.\ref{iota0}.  Again the agreement is extremely good showing that
the collapse depth is controlled by $\iota(0)$, which plays, for small
enough values, the same role of density fluctuations.
\begin{figure}
\vspace{1cm} { \par\centering
  \resizebox*{9cm}{8cm}{\includegraphics*{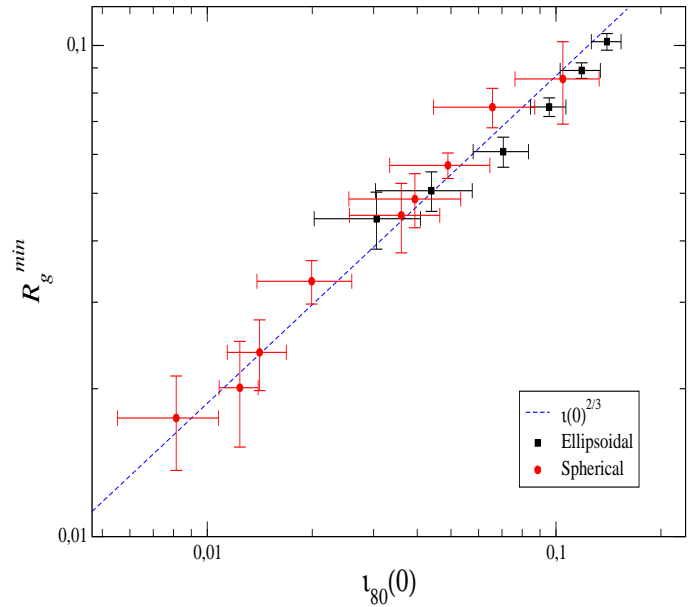}}
  \par\centering }
\caption{Average over 20 realizations of the minimum of the
  gravitational radius vs $\iota(0)$ (black dots).  The case of a
  uniform sphere with different $N$ is also shown (red dots).  The
  dashed line represents the fit with Eq.\ref{Rg-iota}.}
	\label{fig-R_min}
\end{figure}
  
\end{itemize}

\begin{figure}
\vspace{1cm} {\par\centering
\resizebox*{9cm}{8cm}{\includegraphics*{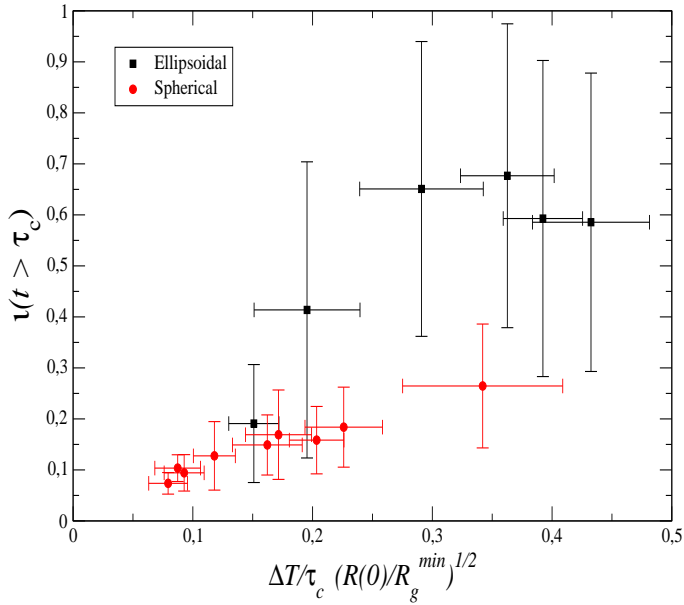}}
\par\centering }
\caption{Evolution of the asymptotic value of $\iota$ as a function of
  $\Delta T \times \sqrt{R_g^{min}}^{-1}$, for both ellipsoidal and
  spherical initial conditions. 
  Eq.\eqref{iota-estimation}} 
\label{figDeltaT_vc}
\end{figure}

{The asymptotic value of $\iota_{80\%}$ as a function of $\Delta
  T \times R_g^{min}$, for different initial flattening ratios and
  averaged over 20 realizations, is shown in Fig. \ref{figDeltaT_vc}.}
We note that the linear behavior described by Eq.\ref{iota-estimation}
nicely fit the observed behavior for the spherical simulations.
However for the ellipsoidal case, when $\iota(0)>0.1$, the asymptotic
value of $\iota_{80}$ clearly deviates from the simple growing
behavior.  This occurs because of the formation of substructures, a
situation that clearly becomes more complex than the one we considered
above.

In summary the violent relaxation mechanism as illustrated by
Figs.\ref{fig-DeltaT_vs_iota0}-\ref{fig-R_min} applies equally well to
both the spherical and the ellipsoidal initial conditions, considering
that finite $N$ density fluctuations in the former case play the same
role of the ratio between the largest and the smallest semi-axis in
the latter case.
%


\section{Virialized and ejected particles} 
\label{VEP} 

We first discuss the properties of the virialized structure formed
after the gravitational collapse, {highlighting that { some of
    their main statistical properties}, such as the density and
  velocities profiles, are formed during the violent relaxation
  mechanism and do not show an imprint of the initial conditions. This
  result, of course, does not hold in general for any cold initial
  density profile: the identification of the phase space parameters of
  {a} cold initial cloud for which {the same} density and
  velocity profiles are formed goes beyond the scope of the present
  work.}  Then, we describe the different configurations that ejected
particles can form, which instead depend on the initial cloud shape.


\subsection{The universal density and velocity profiles
 of bound particles} 

\begin{figure}
\vspace{1cm} { \par\centering
  \resizebox*{9cm}{8cm}{\includegraphics*{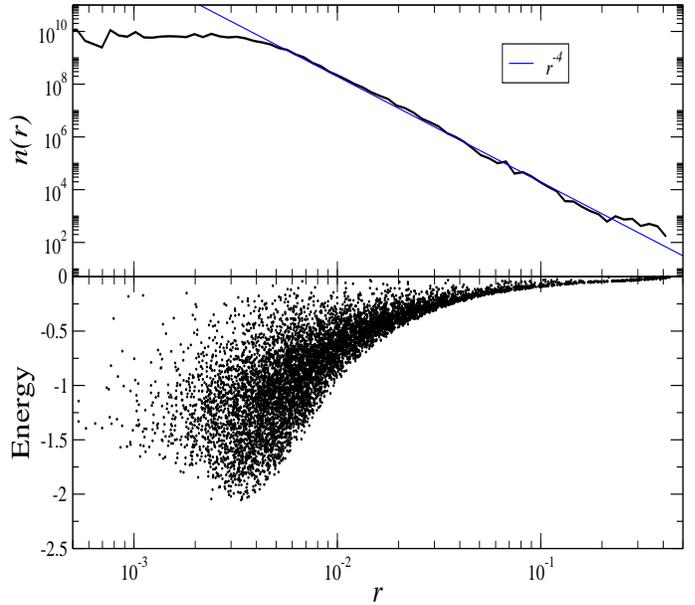}}
  \par\centering }
\caption{Upper panel: Density profile for the virialized state formed
  in the collapse of a spherical cloud (a line with slope $r^{-4}$ is
  plotted for reference).  Bottom panel: particle energy (for bound
  particles only and in normalized units) as a function of the
  distance from the center of mass.  }
\label{energy-dp}
\end{figure}

\begin{figure}
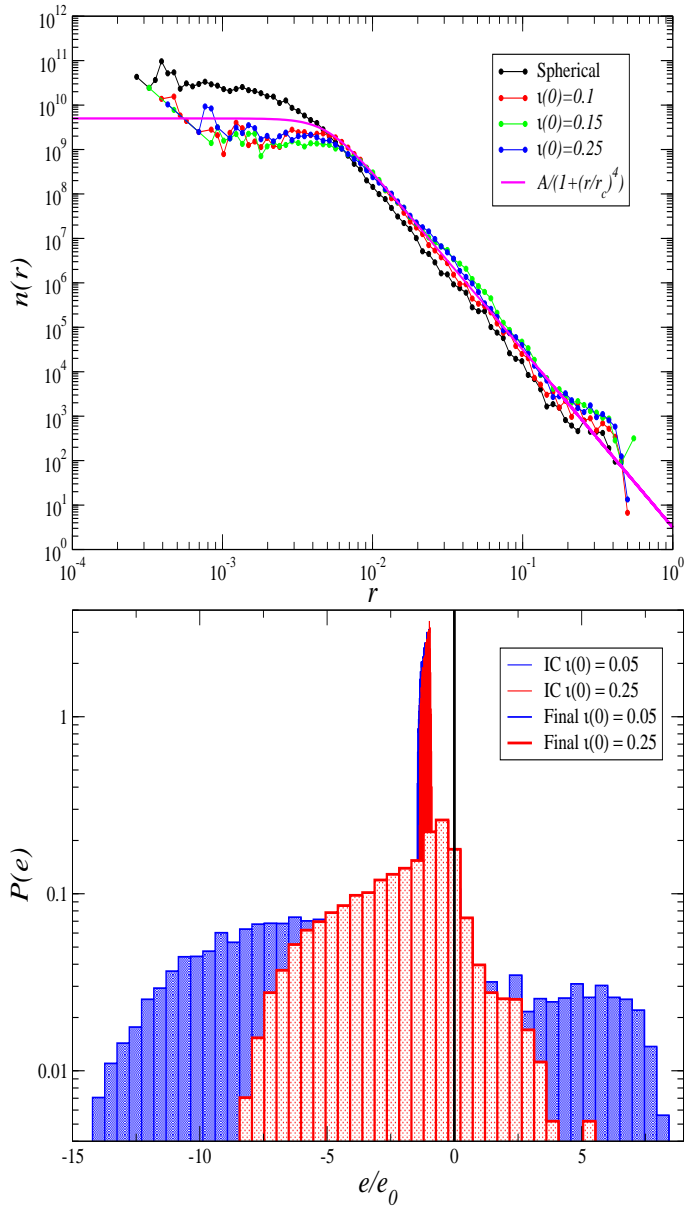

\vspace{1cm} { \par\centering
  \resizebox*{9cm}{8cm}{\includegraphics*{Fig14.eps}}
  \resizebox*{9cm}{8cm}{\includegraphics*{Fig14b.eps}}
  \par\centering }
\caption{Upper Panel: Density profile for clouds with different
  initial shape: spherical and prolate ellipsoidal with $\iota(0)$ as
  reported in the labels.  As a reference we have { also reported
  } the behavior of the quasi equilibrium profile.  Lower panel:
  particle distribution energy at both the initial and the final time
  for ellipsoidal initial conditions respectively with $\iota(0)=0.05$
  and $\iota(0)=0.25$.  {The particle energy $e$ has been normalized
    to $e_0$, the absolute value of the the potential energy of a
    particle placed at distance $R_c$ from the center of a uniform
    spherical cloud.}}
\label{profile}
\end{figure}

\begin{figure}
\vspace{1cm} { \par\centering
  \resizebox*{9cm}{8cm}{\includegraphics*{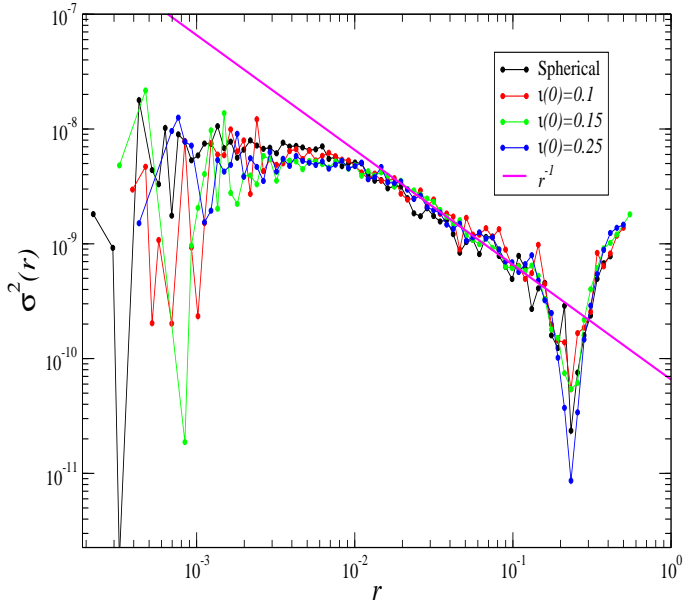}}
  \par\centering }
\caption{Radial velocity dispersion for clouds with different initial shape.
As a reference we have reported also the
Keplerian $1/r$ behavior.} 
\label{sigma2_ell}
\end{figure}

As mentioned in the introduction, it was recently shown that the
post-collapse virialized structure, formed when a cloud is initially
cold enough, has a universal density profile. That is, the same
profile was found when the initial density profile was uniform, power
law with $0 \le \alpha \le 2$, Plummer, Gaussian or Hernquist
\citep{Syloslabini_2012b}. The best fit density profile is 
\be
\label{qep}
\rho(r) = \frac{\rho_0}{1+ (r/r_0)^4} 
\ee 
where, in the uniform case, $\rho_0$ and $r_0$ depend on the number of
points $N$.  This profile was shown to form when bound particles can
have energies close to zero, so that they are allowed to orbit
{mostly} in radial trajectories around the dense core, forming
thus the $\rho(r) \sim r^{-4}$ tail (see Fig.\ref{energy-dp}).  We
stress that this situation occurs when a fraction of the system
particles get positive energy after the collapse.  Thus {the mass
  and energy} ejection and the formation the { $\rho(r) \sim
  r^{-4}$} profile are two aspects of the same phenomenon intrinsic to
the violent relaxation mechanism.
Fig.\ref{profile} shows that Eq.\ref{qep} {gives an} excellent
fit also when the initial condition is an ellipsoid; this fact
suggests a common origin for the density profile as an outcome of the
violent relaxation mechanism. In addition, the radial velocity
dispersion shows a Keplerian behavior for all cases (see
Fig.\ref{sigma2_ell}) which is a further key element in the physical
model for the formation of the $\sim r^{-4}$ tail.
 {This means that the initially ellipsoidal cloud, in the range of
   parameter space considered here, {shows the same gravitational
     collapse dynamics as the spherical clouds.} In particular, the
   virialized structure emerging from the violent relaxation share the
   same density and velocity profiles. As argued in
   \cite{Syloslabini_2012a,Syloslabini_2012b} these are formed during
   the violent relaxation mechanism because of the large energy
   changes which particles undergo during the collapse.}

\begin{figure}
\vspace{1cm} { \par\centering
  \resizebox*{9cm}{8cm}{\includegraphics*{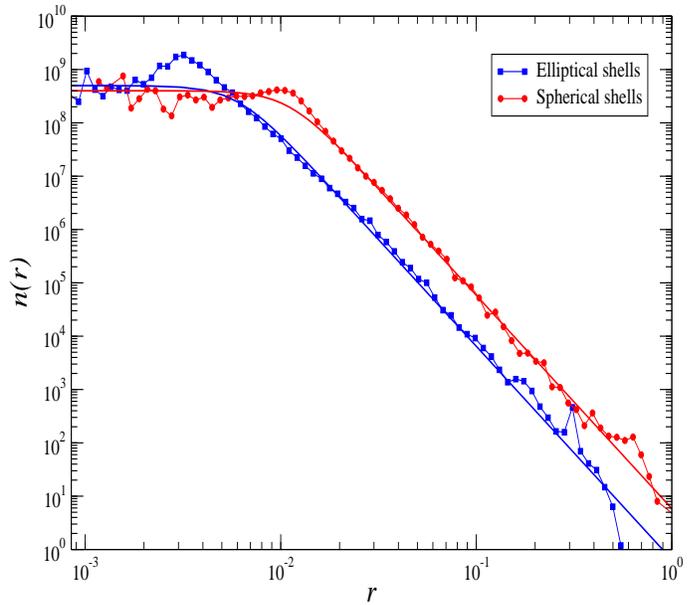}}
  \par\centering }
\caption{Profiles calculated with ellipsoidal shells and spherical shells,
for the case $\iota(0)=0.15$.
} 
\label{ell_shells}
\end{figure}

  {As noticed above, the remnants of cold collapses of initially
    ellipsoidal clouds are strongly non-spherical.  In order to take
    into account the elongated cloud shape, ellipsoidal shells of
    particles are used in the density profile calculation: in
    particular, we computed the density profile in ellipsoidal shells
    with parameters proportional to (and smaller of) those of the
    final structure.  In Fig.\ref{ell_shells} we show density profiles
    and their respective fits, using elliptical and spherical shells
    for the case $\iota(0)=0.15$.  While the free parameters $\rho_0$
    and $r_0$ are different for these two profiles, one can see that
    Eq.\ref{qep} still provides an excellent fit to the density
    profile.  }

  {The distinctive feature of the violent relaxation mechanism is
    represented by the large particle energy change. This is
    illustrated in Fig.\ref{profile} (lower panel), where it is
    plotted the initial and final particle energy distribution $P(e)$
    for $\iota(0)=0.05$ and $\iota(0)=0.25$.  { As the collapse
      is softer in the former case,} the final $P(e)$ has a smaller
    spread than for case $\iota(0)=0.05$: in particular, both the
    small and high energy tails are reduced. This corresponds to the
    fact that the final core is less dense and there are fewer ejected
    particles.}

\subsection{The asymmetric feature of particle ejection}


%

\begin{figure}
\vspace{1cm} { \par\centering
  \resizebox*{9cm}{7cm}{\includegraphics*{Fig17a.eps}}
  \resizebox*{9.5cm}{6.65cm}{\includegraphics*{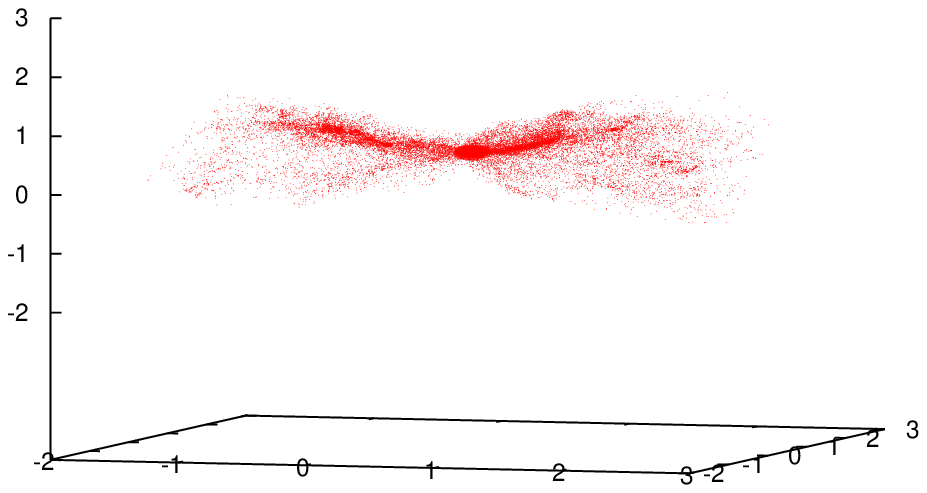}}
  \resizebox*{9.5cm}{6.65cm}{\includegraphics*{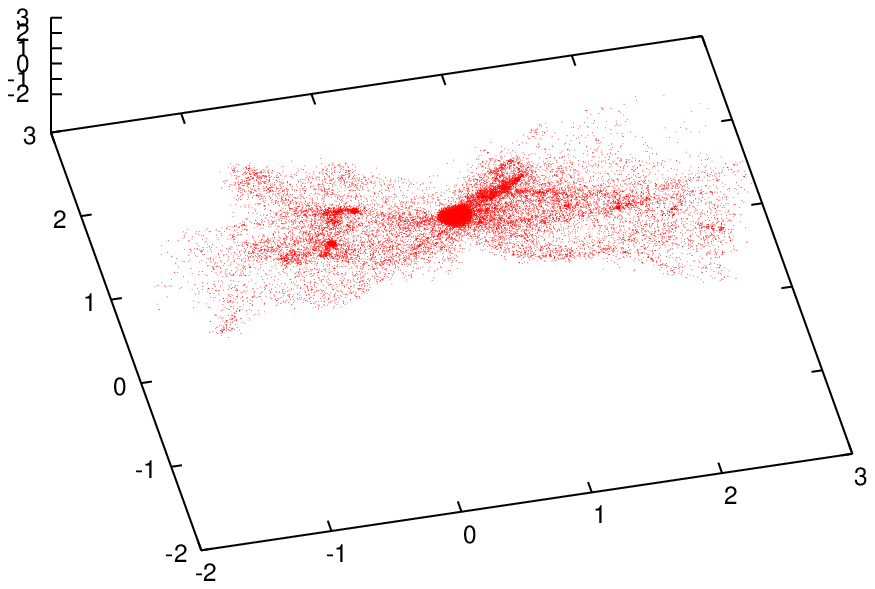}}
             \par\centering }
\caption{Upper Panel: $\iota,\tau, \phi$ of ejected particles for a
  prolate initial ellipsoidal cloud as a function of the initial
  $\iota(0)$. Middle and lower panels: Projections onto two different
  planes of both ejected and bound particles (that are clustered in
  the center) for prolate initial condition with $N=10^5$ and
  $\iota(0)=0.25$. The bar-like shape of ejected particles is clearly
  visible.}
\label{prolate_projection}
\end{figure}

\begin{figure}
\vspace{1cm} { \par\centering
  \resizebox*{9cm}{7cm}{\includegraphics*{Fig18a.eps}}
  \resizebox*{10cm}{7cm}{\includegraphics*{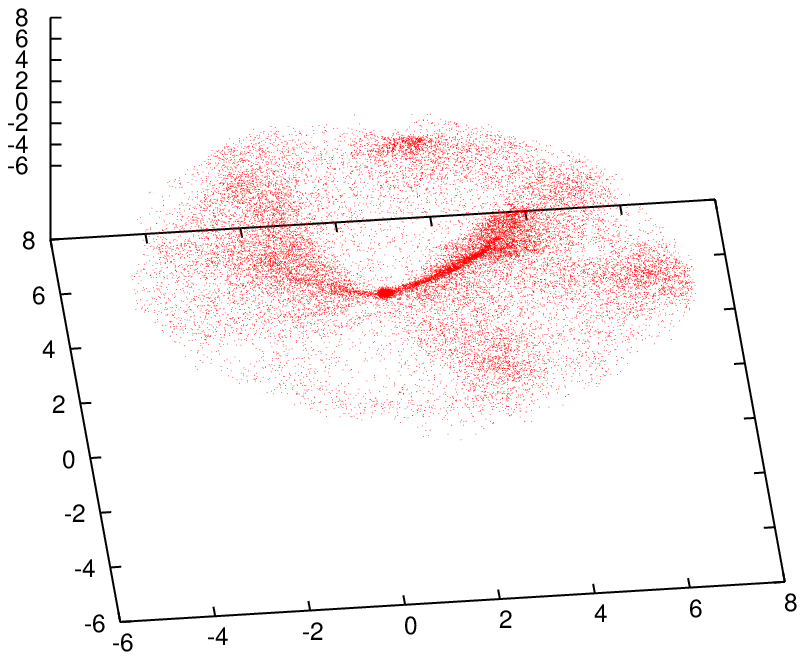}}
  \resizebox*{10cm}{7cm}{\includegraphics*{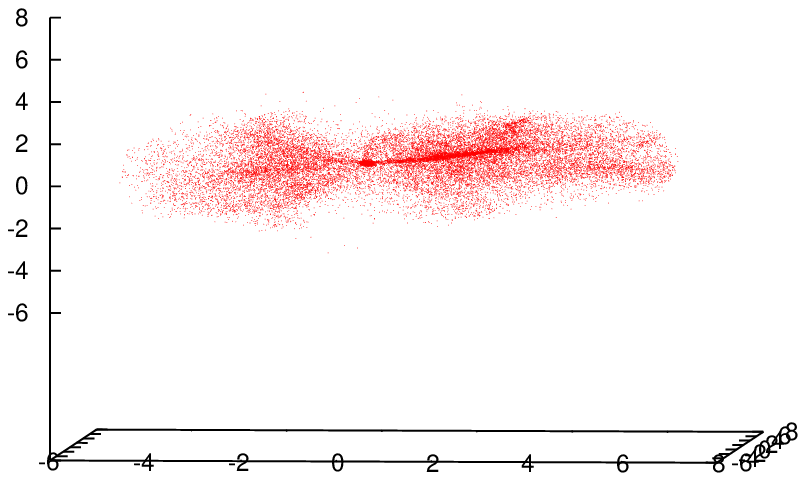}}
             \par\centering }
\caption{As Fig.\ref{prolate_projection} but for oblate initial clouds
  with $N=10^5$ and $\iota(0)=0.05$. }
\label{oblate_projection}
\end{figure}

\begin{figure}
\vspace{1cm} { \par\centering
  \resizebox*{9cm}{7cm}{\includegraphics*{Fig19a.eps}}
  \resizebox*{10cm}{7cm}{\includegraphics*{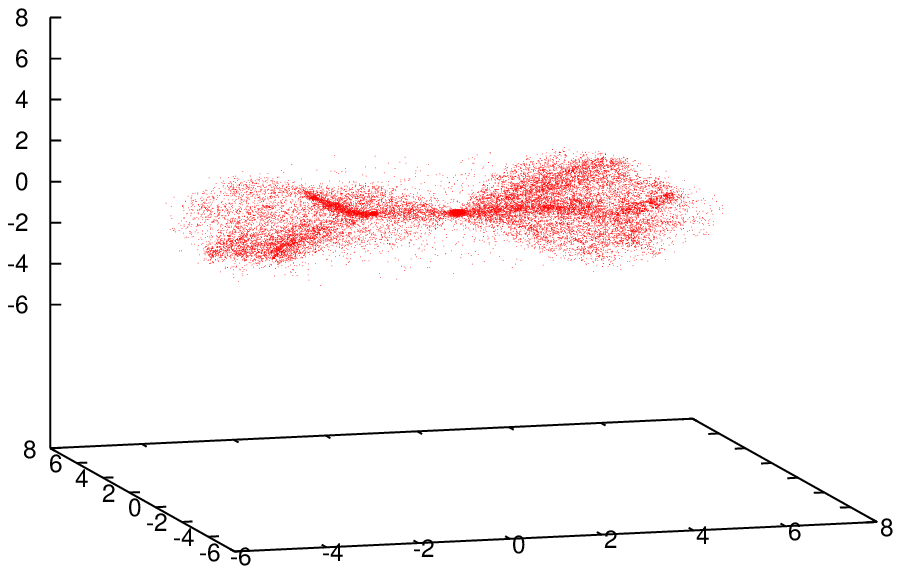}}
  \resizebox*{10cm}{7cm}{\includegraphics*{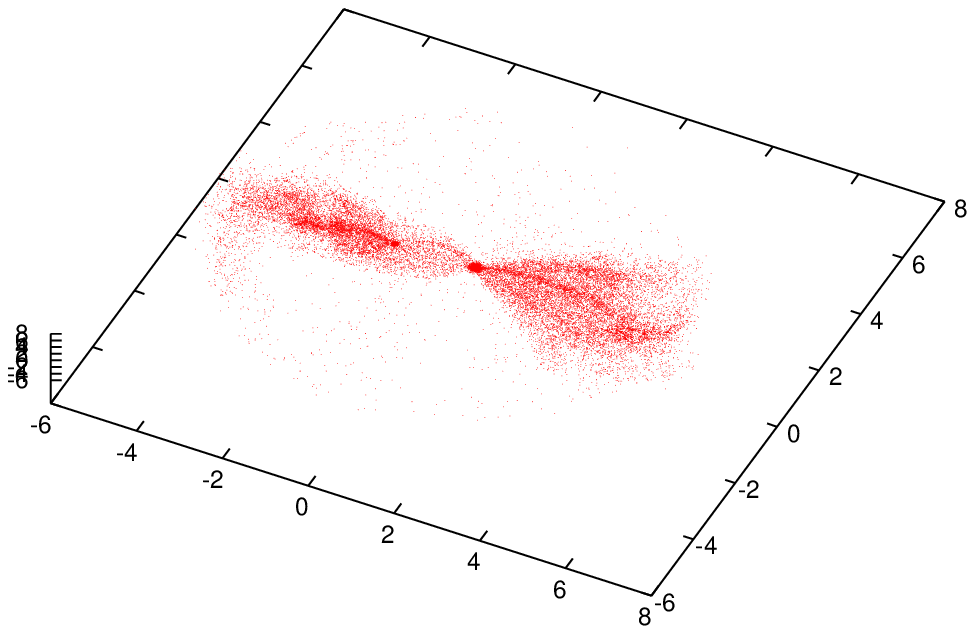}}
             \par\centering }
\caption{As Fig.\ref{prolate_projection} but for triaxial initial
  clouds with $N=10^5$ with axis $a_3=1.1$, $a_2=1.05$, $a_1=1$.}
\label{triaxial_projection}
\end{figure}

As discussed above, because ejected particles are those initially at
$a_1=a_2 \le r \le a_3$, they are not spherically symmetric neither
initially {nor} for $t> \tau_c$: indeed, the collapse amplifies
their initial asymmetry resulting in a very anisotropic final
distribution. To characterize their shapes we measure the $\iota,
\tau$ and $\phi$ (see the upper panel of
Fig.\ref{prolate_projection}): we may see that by increasing
$\iota(0)$, ejected particles do not maintain spherical symmetry (as
for $\iota(0)=0$).  Instead, they first tend to form a flat structure
and then a bar for $\iota(0) > 0.1$ (see
Fig.\ref{prolate_projection}). Obviously particles ejected on the two
opposite sides of the virialized structure will have velocities
oriented in opposite directions, i.e. they will be anti-correlated.
When the initial ellipsoid is oblate then ejected particles form
a triaxial structure.  Indeed, Fig.\ref{oblate_projection} (upper
panel) shows that $\iota \approx 2 \phi$ and $\tau \approx 1/2$.
%
Finally for a triaxial initial cloud ejected particles form an
anisotropic structure that clearly depends on the choice of the initial
axes (see Fig.\ref{triaxial_projection}).

 As we can see in Figs. \ref{prolate_projection},
 \ref{triaxial_projection} and \ref{oblate_projection} by changing the
 value of the initial flatness ratio $\iota(0)$ and the initial shape
 of the structure (i.e. prolate, oblate or triaxial) one can form
 different final distributions of ejected particles which are close to
 flat.  While in most cases the majority of the free particles are
 ejected within a bar along the major semi-axis of the final
 structure, we note that for the oblate case, especially when
 $\iota(0)=0.05,0.1$, the distribution of the ejected particles is
 closer to that of a disk.


\section{Discussion and conclusions} 
\label{DC} 

The gravitational collapse of an initially cold and ellipsoidal cloud
of particles shows non-trivial characteristics: some of them are
similar to the ones observed {when an initially spherical cloud
  is considered {as an initial condition}. In particular, we have
  focused {our attention} on the the formation of power-law
  density and radial velocity profiles and on the ejection of a
  fraction of mass and energy from the system}. The analogy between
the spherical and the ellipsoidal cloud collapse can be understood
considering that the initial deviations from spherical symmetry in the
former case plays the same dynamical role of density fluctuations in
the latter one.

Indeed, we have shown that in the ellipsoidal case the ratio between
the largest $a_3$ and smallest $a_1$ semi-axis play the same role of
density fluctuations for a spherical cloud. {Indeed,} $N$
randomly placed particles in a sphere, have an intrinsic ellipsoidal
deformation $\propto \sqrt{N}^{-1}${ due to Poisson fluctuations;
  thus} increasing $N$ has the same effect of decreasing the
difference between $a_3$ and $a_1$.
We have discussed in detail that the mechanism of particles energy
gain during the collapse gives rise to the ejection of those particles
that are initially placed close to the cloud boundaries: for an
ellipsoidal cloud this implies an asymmetric ejection.

Thus, the virialized state formed after the collapse {of an
  ellipsoidal cloud} shows { the same features of the cold spherical
  cloud remnant: in both cases the density and radial velocity
  profiles arise from the violent relaxation mechanism. Spherical
  clouds with different density profiles can however form different
  virialized structures and a more complete study of the cloud
  parameters phase space is necessary to draw more general conclusions
  about quasi-equilibrium profiles. }

  Note that the mechanism described here is, in principle, different
  from the the radial orbit instability discussed by
  \cite{Antonov_1961,Fridman_Polyachenko_1984}. This instability
  characterizes a different case from the one considered here, i.e. a
  spherically symmetric stationary solutions of the collisionless
  Boltzmann equation with purely radial orbits.  {The radial
    orbits instability characterizing equilibrium models and systems
    undergoing to strong collapses and violent relaxations are
    distinct, but probably connected, physical mechanisms. Indeed, as
    firstly noticed by \citet{Merritt_Aguilar_1985} a radial
    instability can be the mechanism that, in the collapse of an
    initially cold system in which particles have mainly radial
    orbits, gives rise to a triaxial structure.  Our findings clarify
    the dynamics driving the formation of a bar/disk structure in more
    detail than in previous works, by focusing on the relation between
    initial and final conditions for simple non-spherical systems.
    However a detailed theoretical understanding of the radial orbit
    instability for non-equilibrium collapsing systems is still an
    open problem: we refer the interested reader to a forthcoming work
    for a more detailed discussion about this point
    \citep{Syloslabini_2014}.  }

On the other hand ejected particles form non-trivial
anisotropic configurations, like bars or disks, that depend on the
initial cloud shape: a small initial deviation from spherical symmetry
is amplified by the violent relaxation energy gain mechanism and
gives rise to highly non-spherical clouds. The ellipsoidal clouds
that we have considered represent thus a further step toward a more
realistic cloud collapse, that has allowed us to perform controlled
N-body experiments to quantify the effect of the initial sphericity.

 { This result, although obtained for a very simplified and
   unrealistic physical cloud model, is interesting as it points out
   that it is relatively natural to generate, from an initial
   condition close to isotropic, a central (host) structure surrounded
   by particles, or even clusters of particles, that are strongly
   anisotropically distributed, even sometimes close to a planar
   distribution.  {Galaxy formation is a much more complex
     process, involving physical mechanisms such as gas dynamics,
     effects of dark matter halos, etc.,} which are not taken into
   account in this work. { However, it} is worth mentioning that
   recently several observational studies have pointed out that dwarf
   satellites of galaxies both the Milky Way and the Andromeda galaxy
   are non isotropically distributed around their host galaxy (see
   \cite{Ibata_2013,Ibata_2014} and references therein).  In
   particular it was observed that dwarf satellites are aligned in a
   thin plan structure and that they have coherent kinematic
   properties.  {Our results suggest that these} particular
   features might be better understood from the collapse of a
   non-spherical cloud.  In a forthcoming work we will consider more
   realistic initial conditions and a detailed comparison with
   observations.  }
\bigskip

We thank Michael Joyce for useful discussions and comments. {We
  also thank an anonymous referee for a number of interesting comments
  that have allowed us to improve the presentation of our work}.
Numerical simulations have been run on the Cineca supercomputer
(project ISCRA QSS-SSG) and on the super-computer “Mesu” of UPMC.

\bibliographystyle{mn2e}


\end{document}